\documentclass[aps,prb,twocolumn,groupedaddress,showpacs]{revtex4}

\usepackage{graphicx}
\usepackage{latexsym}
\usepackage{amsmath,amssymb}
\usepackage{bm}

\newcommand{\bnabla}{\bm{\nabla}}

\begin{document}

\title{
Metal-Insulator transition in Hubbard-like models with random 
hopping}

\author{Matthew S. Foster}
\email{psiborf@physics.ucsb.edu} 
\author{Andreas W. W. Ludwig}

\affiliation{Department of Physics, University of California, Santa
Barbara, CA 93106}

\date{\today}

\begin{abstract} 
An instability of a diffusive Fermi liquid, indicative of a metal-insulator 
transition (expected to be of first order), arising solely from the competition 
between quenched disorder and short-ranged interparticle interactions is 
identified in Hubbard-like models for spinless fermions, subject to (complex) 
random hopping at half-filling on bipartite lattices. The instability,
found within a Finkel'stein non-linear sigma model treatment in
$d=(2+\epsilon)>2$ dimensions, originates from an underlying 
particle-hole like (so-called chiral) symmetry, shared by both disorder 
and interactions. In the clean, interacting Fermi liquid this symmetry is 
responsible for the (completely different) nesting instability.
\end{abstract}


\maketitle

Understanding the combined effects of quenched disorder and
interparticle interactions in electronic systems remains one of 
the central problems in solid state physics. Models of noninteracting 
electrons subject to static, random impurity potentials provide the simplest 
description of disordered metals; many analytical and numerical studies have shown that 
such models exhibit a continuous metal-insulator transition (MIT) in three dimensions 
$(3\textrm{D})$,\cite{LeeRamakrishnan} accessed by varying either the disorder 
strength, or the Fermi energy relative to the mobility edge. By contrast, all 
electronic states are typically exponentially localized in one and two dimensions 
by arbitrarily weak disorder. (Notable exceptions occur in the presence of spin-orbit 
scattering, and in the systems discussed, e.g., in Refs.~\onlinecite{1DChiral,Gade}.)

The above description ignores the important effects of interparticle interactions.
Unfortunately, however, theories capturing the competition of \emph{both} 
disorder and interactions are typically quite complex, and difficult to analyze 
reliably.\cite{Finkelstein,BK} These issues have come again to the forefront of scientific debate 
in view of discussions concerning a metal-insulator transition in $2\textrm{D}$, and 
related experimental results on $2\textrm{D}$ semiconductor inversion layers.\cite{2DMIT}

\begin{figure}
\includegraphics[width=0.4\textwidth]{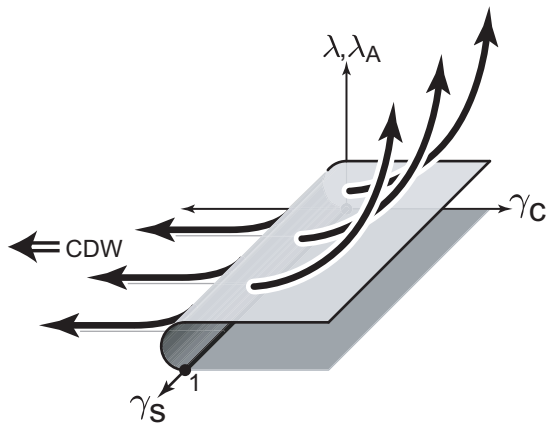}
\caption{Schematic phase diagram in $d = 2 + \epsilon$, with $0 < \epsilon \ll 1$. 
$\lambda$ and $\lambda_A$ are measures of the disorder, whereas $\gamma_s$ and $\gamma_c$ 
characterize the interaction strengths. The stable metallic phase resides 
between the ballistic plane ($\lambda = \lambda_{A} = 0$) and the shaded sheath; 
the thick arrows indicate the two instabilities of the (metallic) diffusive Fermi liquid 
discussed in the text.
\label{PhaseDiag}}
\end{figure}

In the present work, we identify a novel ``Anderson-Mott'' instability of a diffusive 
Fermi liquid in 
\linebreak
$d=(2+\epsilon)>2$ spatial dimensions ($\epsilon \ll 1$),
which arises solely from the competition between disorder and short-range 
interactions. This instability is indicative of a metal-insulator transition (expected
to be of first order) from the diffusive Fermi liquid to an insulating state dominated by 
both strong disorder and interactions. (See Fig.~\ref{PhaseDiag}, below.) Since the system 
that we study has no localized phase with disorder in the absence of interactions, a 
localized phase can only appear due to the presence of the interactions. We expect our result 
to be relevant for sufficiently strong disorder in $d = 3$ dimensions.

Specifically, we analyze a class of ``Hubbard-like'' models\cite{Auerbach} of
spinless fermions, at half-filling on bipartite lattices, with random (short-range) hopping 
between the two sublattices. [The Hamiltonian is given below in Eq.~(\ref{H}).] 
In every realization of the (static) disorder, such a model possesses a special 
particle-hole like symmetry, which we will refer to as {\it sublattice symmetry} (SLS). 
[SLS is termed ``chiral symmetry'' in the classification scheme of Ref.~\onlinecite{Zirnbauer} 
(see also Refs.~\onlinecite{1DChiral,Gade,ChiralOne,GLL,FC,ChiralTwo,BDIpaper}).]

In the absence of disorder, it is SLS which is responsible for the ``nesting'' condition of 
the Fermi surface. Fermi surface nesting is, in a sense, the defining property of 
(clean) Hubbard-like models for interacting lattice fermions in $d \geq 2$. It is the
nesting condition which makes the ballistic Fermi liquid phase at half filling in such models 
unstable to Mott insulating order in the presence of generic, arbitrarily weak interparticle 
interactions.\cite{Hirsch,GSST,Shankar} Here we consider, in addition, 
complex random (nearest-neighbor) hopping, which breaks time reversal invariance (TRI) in 
every realization of disorder.\cite{ClassAIII} For our system of spinless fermions, this is 
consistent with the application of a random magnetic field to the otherwise clean model. 

Our principal motivation for studying such a model is that we expect both disorder 
and interparticle interactions to play important roles in the description of the low-energy 
physics. Because random hopping preserves the special SLS, our disordered model retains the 
nesting instability of the associated clean system. This instability can therefore compete with 
the unusual localization physics of the disordered, but noninteracting model (see below). 
The further assumption of broken TRI guarantees that we do not have to confront an additional 
superconducting instability.\cite{BK,Hirsch,GSST,Shankar} We expect the instability that 
we identify in this work in the simultaneous presence of both disorder and interactions to occur in
three dimensions for sufficiently strong disorder, and we stress that it is clearly distinct from 
the pure Mott nesting instability, although the latter also appears in our model 
phase diagram (see Fig.~\ref{PhaseDiag}, below). We note that the effects of hopping disorder 
upon the N\'eel ground state of the (slightly more complex) spin-$1/2$ Hubbard model at half 
filling were studied numerically in Ref.~\onlinecite{Scalettar}, although these 
studies were limited to $d = 2$. 

A second motivating factor is that, interestingly, the presence of SLS radically 
changes the localization physics of the disordered, noninteracting random 
hopping model [Eq.~(\ref{H}), below, with $V = U = 0$]. SLS enables the random hopping 
model to evade the phenomenon of Anderson localization. Specifically,
the noninteracting system exhibits a critical, delocalized phase at the band center 
(half filling) in one, two, and three dimensions for finite disorder strength, with a 
strongly divergent low-energy density of states in 
$d =1,2$.\cite{1DChiral,Gade,ChiralOne,GLL,FC,ChiralTwo}
In particular, there is no MIT and no Anderson insulating phase in $d = 3$ 
(in the absence of interactions). Random hopping models have been of significant theoretical 
interest in the recent past, both because of the unusual delocalization physics described above, 
but also because these models have proven amenable to a variety of powerful analytical 
techniques in $d \leq 2$, with many exact and/or non-perturbative features 
now understood.\cite{1DChiral,GLL,ChiralTwo} This situation should be contrasted with 
our understanding of the conventional noninteracting (``Wigner-Dyson'') MIT, 
which is based largely on perturbative results in $d > 2$.\cite{LeeRamakrishnan} 
Our work here addresses the effects of interparticle interactions in random 
hopping models for $d\geq 2$.

Our starting point is the following extended Hubbard-like Hamiltonian for 
spinless fermions hopping on a bipartite lattice at half filling:
\begin{align}\label{H}
	H =& - \sum_{\langle i j \rangle}\left(t + \delta t_{i, j}\right)
	c_{A i}^{\dagger} c_{B j}^{\phantom{\dagger}} + \mathrm{H.c.} 
	+ 2 V \sum_{\langle i j \rangle} 
	\delta\hat{n}_{A i}
	\delta\hat{n}_{B j} \nonumber \\
	& + U \bigl (
	\sum_{\langle\langle i i' \rangle\rangle} 
	\delta\hat{n}_{A i} \delta\hat{n}_{A i'} 
	  + 
	\sum_{\langle\langle j j' \rangle\rangle} 
	\delta\hat{n}_{B j} \delta\hat{n}_{B j'}\bigr ).
\end{align}
Any bipartite lattice may be divided into two interpenetrating sublattices, 
which we distinguish with the labels $A$ and $B$.
In Eq.~(\ref{H}), $c_{A i}^{\dagger}$ and $c_{B j}^{\phantom{\dagger}}$ denote 
fermion creation and annihilation operators on the $A$ and $B$ sublattices, respectively. 
Here, $i$ and $j$ respectively index the $A$ and $B$ sublattice sites, and the sums on 
$\langle i j \rangle$ run over all nearest neighbor $A$-$B$ lattice bonds, while the sums 
on $\langle\langle i i' \rangle\rangle$ and $\langle\langle j j' \rangle\rangle$ 
run over all next-nearest neighbor (same sublattice) pairs of sites. The homogeneous 
hopping amplitude $t$ in Eq.~(\ref{H}) is taken to be purely real; disorder 
appears in the perturbation $\delta t_{i, j}$. We take the amplitude 
$\delta t_{i, j}$ to be a Gaussian complex random variable 
with zero mean, statistically independent on different lattice links. The operators 
$\delta\hat{n}_{A/B} \equiv (c_{A/B}^{\dagger} 
c_{A/B}^{\phantom{\dagger}} - \frac{1}{2})$ in Eq.~(\ref{H}) denote the deviations 
of the local sublattice fermion densities from their value at 
half filling; the interaction strengths $V$ and $U$ appearing in this equation 
couple to nearest neighbor and next-nearest neighbor density-density interactions, 
respectively.

The Hamiltonian $H$ in Eq.~(\ref{H}) is invariant under the (antiunitary) sublattice 
symmetry (SLS) transformation
\begin{equation}\label{SublatticeSym}
	c_{A i}^{\phantom{\dagger}} \rightarrow c_{A i}^{\dagger},
	\;\; c_{B j}^{\phantom{\dagger}} \rightarrow - c_{B j}^{\dagger},
\end{equation}
(all complex scalar terms in $H$ are complex conjugated). \cite{ClassBDI}
In the clean limit, Eq.~(\ref{H}) with $\delta t_{i, j} = 0$ for all
lattice bonds
$\langle i j \rangle$,  SLS is responsible for the nesting condition of the 
noninteracting Fermi surface. As a result of nesting, the Fermi liquid phase of 
the clean model is unstable to charge density wave (CDW) order for any 
$2 V > U \geq 0$.\cite{Shankar,AIIIlongpaper}

The effects of the interparticle interactions $U$ and $V$ upon the delocalized phase 
of the disordered model given by Eq.~(\ref{H}) may be investigated by using 
Finkel'stein's generalized non-linear sigma model (FNL$\sigma$M) approach\cite{Finkelstein,BK}
to formulate the low-energy effective continuum field theory. The latter can be studied
using a controlled $\epsilon$ expansion in $d = 2 + \epsilon$ dimensions
($0 \leq \epsilon \ll 1$). We use the Schwinger-Keldysh\cite{KELDYSH} method to perform 
the ensemble average over realizations of the hopping disorder. The FNL$\sigma$M is derived 
following the standard methodology.\cite{Finkelstein,BK,KELDYSH}

In the present case, the resulting FNL$\sigma$M is described by the generating 
functional\cite{AIIIlongpaper}
\begin{equation}\label{Z}
	Z = \int \mathcal{D} \hat{Q} \; e^{-S_{D}-S_{I}},
\end{equation}
where
\begin{align}
	S_{D} = 
	\frac{1}{8 \pi \lambda} & \int d^{d}{\bm{\mathrm{r}}} \,
	\mathrm{Tr}
	\left[
	\bnabla\hat{Q}^{\dagger}({\bm{\mathrm{r}}}) 
	\cdot \bnabla\hat{Q}({\bm{\mathrm{r}}}) 
	\right]
	\nonumber \\
	+ h & \int d^{d}{\bm{\mathrm{r}}} \,
	\mathrm{Tr}
	\left\{
	i \hat{\Omega}
	[\hat{Q}^{\dagger}({\bm{\mathrm{r}}})+\hat{Q}({\bm{\mathrm{r}}})]
	\right\}
	\nonumber \\
	- & \frac{\lambda_{A}}{8 \pi \lambda^{2}} \int d^{d}{\bm{\mathrm{r}}} 
	\left\{
	\mathrm{Tr}
	\left[\hat{Q}^{\dagger}({\bm{\mathrm{r}}})
	\bnabla\hat{Q}({\bm{\mathrm{r}}})\right]
	\right\}^{2}
	, \label{SD}
\end{align}
and
\begin{align}
	S_{I} = & i \sum_{a=1,2} \xi^{a} 
	\int dt \, d^{d}{\bm{\mathrm{r}}} 
	\left\{
	2 \Gamma_{V} \, {Q}^{\dagger \, a a}_{t t}({\bm{\mathrm{r}}}) 
	{Q}^{a a}_{t t}({\bm{\mathrm{r}}})
	\right. 
	\nonumber \\
	& \left.
	+ \Gamma_{U}[{Q}^{a a}_{t t}({\bm{\mathrm{r}}}){Q}^{a a}_{t t}({\bm{\mathrm{r}}}) 
	+ {Q}^{\dagger \, a a}_{t t}({\bm{\mathrm{r}}})
	{Q}^{\dagger \, a a}_{t t}({\bm{\mathrm{r}}})]
	\right\}
	\label{SI}.
\end{align}
The field variable 
\begin{equation}\label{QmatrixDef}
	\hat{Q}({\bm{\mathrm{r}}}) \rightarrow Q^{a b}_{t t'}({\bm{\mathrm{r}}})
\end{equation} 
in Eqs.~(\ref{Z})--(\ref{SI}) is a complex, ``infinite-dimensional'' square matrix 
living in $d$ spatial dimensions, where indices $t$ and $t'$ belong to a 
continuous time or (via Fourier transform) frequency space, 
and where indices $a$ and $b$ belong to a 2-dimensional ``Keldysh'' species space, 
with $a,b \in \{1,2\}$.\cite{KELDYSH,OmegaMatrixDef} In Eq.~(\ref{SD}), $\mathrm{Tr}$ 
denotes a matrix trace over time (or frequency) and Keldysh indices. 
$\hat{Q}({\bm{\mathrm{r}}})$ satisfies in addition the unitary constraint
\begin{equation}\label{QmatrixUnitary}
	\hat{Q}^{\dagger}({\bm{\mathrm{r}}}) \hat{Q}({\bm{\mathrm{r}}}) = \hat{1}.
\end{equation}
The matrix $\hat{Q}$ and its adjoint $\hat{Q}^{\dagger}$ may be interpreted
\cite{AIIIlongpaper}
as continuum versions of the \emph{same-sublattice} fermion bilinears
\begin{subequations}\label{QmatrixID}
\begin{align}
	Q^{a b}_{t t'} & \sim c^{a}_{A}(t) c^{\dagger b}_{A}(t'), \\
	Q^{\dagger a b}_{t t'} & \sim c^{a}_{B}(t) c^{\dagger b}_{B}(t').
\end{align} 
\end{subequations}

The action given by Eq.~(\ref{SD}) describes the low-energy diffusive physics of the 
noninteracting random hopping model; a replica version of the noninteracting sigma 
model with action $S_{D}$ was originally studied by Gade and Wegner.\cite{Gade} 
This noninteracting sector of the FNL$\sigma$M contains three coupling constants: 
$\lambda$, $\lambda_{A}$, and $h$. The parameter $1/\lambda$ is proportional to the 
dimensionless dc conductance $g$ of the system (i.e., is related to the disorder strength),
while $\lambda_{A}$ denotes a second measure of disorder, unique to this sublattice 
symmetry class, which strongly influences the single-particle density of states.\cite{Gade,FC,GLL} 
The parameter $\lambda_{A}$ may be simply interpreted as characterizing the strength of 
long-wavelength, quenched random orientational fluctuations in bond strength dimerization 
of the random hopping model.\cite{AIIIlongpaper} Finally, $h$ is a dynamic scale factor, 
which determines the dynamical critical exponent $z$ in Eq.~(\ref{lnhFlow}), below, through 
the condition $d \ln h/dl \equiv 0$.\cite{BK,BDIpaper,AIIIlongpaper}
The interparticle interactions appear in $S_{I}$, defined by Eq.~(\ref{SI}). 
Given Eq.~(\ref{QmatrixID}), we may interpret
$Q^{a a}_{t t}({\bm{\mathrm{r}}})$ and $Q^{\dagger a a}_{t t}({\bm{\mathrm{r}}})$
as continuum local density operators on the $A$ and $B$ sublattices,
respectively. Then the interaction couplings $\Gamma_{V}$ and $\Gamma_{U}$ in
Eq.~(\ref{SI}) describe generic short-ranged intersublattice and same-sublattice
density-density interactions, respectively [compare to Eq.~(\ref{H}), above].
Finally, $\xi^{a} = \pm 1$ in Eq.~(\ref{SI}), for Keldysh species label $a = 1, 2$.

Using a Wilsonian frequency-momentum shell background field methodology,\cite{Finkelstein} 
we have performed a one-loop renormalization group calculation on the model defined by 
Eqs.~(\ref{Z})--(\ref{SI}). The calculation is straight-forward, though rather lengthy; 
the details will be published elsewhere.\cite{AIIIlongpaper} Below we simply state our results. 
In order to do so, it is convenient to introduce the following effective 
interaction couplings
\begin{equation}\label{singletCDW}
	\gamma_{s} \equiv \frac{2}{\pi h} (\Gamma_{U} + \Gamma_{V}),\; 
	\gamma_{c} \equiv \frac{2}{\pi h} (\Gamma_{U} - \Gamma_{V}).
\end{equation}
The interaction strength $\gamma_{s}$ couples to the square of the 
(smooth) local charge density in the continuum theory, while $\gamma_{c}$ 
couples to the square of the sublattice staggered charge density. In accordance 
with the discussion in the paragraph below Eq.~(\ref{SublatticeSym}), we expect 
$\gamma_{c} < 0$ to promote charge density wave (CDW) formation, while $\gamma_{c} > 0$ 
should suppress it.

We find the following one-loop RG flow equations for the couplings 
$\lambda$, $\lambda_{A}$, $\gamma_{s}$, $\gamma_{c}$, and $h$ in 
$d=(2+\epsilon)$ dimensions:
\begin{subequations}\label{FlowEqs}
\begin{eqnarray}
	\frac{d \lambda}{d l} & = & -\epsilon \, \lambda - \lambda^{2}\gamma_{c} 
	\nonumber \\
	& & + 2 \lambda^{2} 
	\left[
	1 + \frac{1-\gamma_{s}}{\gamma_{s}} \ln(1-\gamma_{s})
	\right], \qquad	\label{lambdaFlow}\\
	\frac{d \lambda_{A}}{d l} & = & \epsilon \, \lambda_{A} + \lambda^{2} 
	+ 2 \frac{\lambda_{A}}{\lambda} \frac{d \lambda}{d l}, 
	\label{lambdaAFlow}\\
	\frac{d \gamma_{s}}{d l} & = & \lambda_{A} (1-\gamma_{s})
	\left(\gamma_{s} + 2 \gamma_{c} - 2 \gamma_{s} \gamma_{c} \right) 
	\nonumber \\
	& & - \lambda (1-\gamma_{s}) \left( \gamma_{s} + \gamma_{c} 
	- 2 \gamma_{s} \gamma_{c} \right), \label{gammasFlow}\\
	\frac{d \gamma_{c}}{d l} & = & \lambda_{A} \left(\gamma_{c} 
	+ 2 \gamma_{s}\right) - \lambda \left(\gamma_{s} 
	+ \gamma_{c}\right) 
	\nonumber\\
	& & + \lambda \left[ 2 \gamma_{c} \ln(1 - \gamma_{s}) + \gamma_{s} 
	\gamma_{c}\right] - 2 \gamma_{c}^{2}, \label{gammacFlow}
	\\
	\frac{d \ln h}{d l} & = & (d - z) +  \lambda_{A} + \lambda(\gamma_{c} 
	- \gamma_{s}).	\label{lnhFlow}
\end{eqnarray}
\end{subequations}
Here, $l$ is the logarithm of the spatial length scale. These flow equations are given 
to the lowest non-trivial order in the couplings $\lambda$, $\lambda_A$, and $\gamma_{c}$, but 
contain contributions from $\gamma_{s}$ to \emph{all} orders; Finkel'stein's NL$\sigma$M formulation
provides\cite{BK} a perturbative expansion which is controlled by the (small) dimensionless resistance $\lambda$, 
but does not require the interaction strength $\gamma_s$ to be small. 
Before turning to an analysis of our results, Eqs.~(\ref{lambdaFlow})--(\ref{lnhFlow}), 
we provide interpretations for various key terms appearing in them. 
First, the term in square brackets on the second line of Eq.~(\ref{lambdaFlow}) 
is the usual correction to the dimensionless dc resistance $\lambda$,
arising from the short-ranged interparticle interactions,\cite{Finkelstein,AltshulerAronov,NPCS} 
and corresponds\cite{Aleiner} to coherent backscattering of carriers off of disorder-induced 
Friedel oscillations in the background electronic charge density.\cite{AleinerExplanation}
The last term in Eq.~(\ref{gammacFlow}) drives the CDW instability, which is a remnant of the clean 
Hubbard-like model (recall that in our conventions $\gamma_{c} < 0$ signals this instability). 

Now we analyze our results. In $d = 2$ dimensions, integrating 
Eq.~(\ref{FlowEqs}) for generic initial conditions shows that the critical, 
delocalized phase of the half-filled, noninteracting random hopping
model\cite{Gade} is unstable to the effects of short-ranged interparticle interactions. 
We find that either $\gamma_{c} \rightarrow - \infty$ signaling CDW formation,
or that $\lambda, \lambda_{A} \rightarrow \infty$ and $\gamma_{c} \rightarrow + \infty$, 
indicating a flow toward both strong disorder and strong interactions.
Regardless, we expect the $2\textrm{D}$ interacting, disordered Hubbard model to be an insulator 
at zero temperature. This should be compared to an analogous result\cite{BDIpaper} previously 
obtained for a TRI, interacting random hopping model on the honeycomb lattice. This physics 
is consistent with numerical studies\cite{Scalettar} of the half-filled spin-1/2 Hubbard model in 
$d=2$, which have shown that TRI random hopping disorder preserves the 
compressibility gap of the clean Mott insulator, and that the disordered and interacting 
system shows no signs of metallic behavior.

The situation in $d = (2 + \epsilon) > 2$ dimensions is more interesting.
Upon increasing $\epsilon$ from zero, a narrow, irregularly shaped sliver corresponding to a 
stable metallic, diffusive Fermi liquid state opens up in the four-dimensional 
$(\lambda,\lambda_{A},\gamma_{s},\gamma_{c})$ coupling constant space. The sliver
encloses the line $\lambda = \lambda_{A} = \gamma_{c} = 0$, with $-\infty < \gamma_{s} < 1$, 
the entirety of which is perturbatively accessible because the FNL$\sigma$M does not require 
the interaction strength $\gamma_{s}$ to be small. A highly schematic 3D ``projected'' phase 
diagram is depicted in Fig.~\ref{PhaseDiag}. In this figure, the interaction constants reside 
in the horizontal plane, while the vertical direction schematically represents (both) disorder 
strengths; the shaded sheath is a cartoon for the boundary of the stable metallic region, 
which resides between it and the ballistic ($\lambda = \lambda_{A} = 0$) plane.
The ``height'' of the stable metallic region in the ``disorder'' directions $(\lambda,\lambda_{A})$
is controlled by $\epsilon$, although the precise shape and size of the phase boundary varies with
$\gamma_{s}$, and is difficult to characterize analytically. Over the range of perturbatively
small values of $\gamma_{c}$, the stable Fermi liquid phase resides in the region $\gamma_{c} \gtrsim 0$,
and terminates near $\gamma_{c} = 0$.

The flow equations (\ref{lambdaFlow})--(\ref{gammacFlow}) possess no perturbatively accessible, 
nontrivial RG fixed points for $d > 2$, and thus no continuous metal-insulator transition 
can be identified. However, the two instabilities described above for the $2\textrm{D}$ case 
persist for $d > 2$, and become clearly distinct roads out of the metallic state. 
The conventional CDW instability always occurs for initial $\gamma_{c} < 0$ and sufficiently 
weak disorder, i.e., when $\lambda, \lambda_{A} \ll \epsilon$, and is represented by the flow 
$\gamma_{c} \rightarrow - \infty$. This flow is accompanied by a decay in both disorder 
strengths $\lambda, \lambda_A$. 

The primary result of this paper is the identification of a second route
out of the diffusive Fermi liquid phase in $d = (2 + \epsilon) > 2$ dimensions, 
independent of the Mott CDW instability, arising solely from the competition of disorder 
and interaction effects. As in the $2\textrm{D}$ case, this second route is 
characterized by a flow off to both strong disorder ($\lambda, \lambda_{A} \rightarrow \infty$) 
and strong interactions ($\gamma_{c} \rightarrow +\infty$), as indicated by the 
thick arrows emerging from the $\gamma_{c}>0$ portion of the phase boundary shown in 
Fig.~\ref{PhaseDiag}; we call it an Anderson-Mott instability. 
Even though there is no perturbatively accessible fixed point, this new
instability is nonetheless perturbatively controlled in $d = (2 + \epsilon)$ 
over a wide range of initial conditions when $\epsilon \ll 1$; in particular,
it is perturbatively accessible over the entire range $0 \leq \gamma_{s} < 1$.\cite{NeggammasNote}
Numerically integrating Eqs.~(\ref{lambdaFlow})--(\ref{gammacFlow}) for small 
$\epsilon \ll 1$, we find that the Anderson-Mott instability can apparently 
always be reached by increasing only the dimensionless resistance $\lambda$. 
We expect the boundary separating the flow toward the stable metallic regime
from that toward the regime of the Anderson-Mott instability to represent a 
disorder-driven, first order metal-insulator transition (MIT). We emphasize that 
a MIT does not exist in the noninteracting random hopping model, which possesses 
only a delocalized phase at half-filling for finite disorder in 
$d\geq1$,\cite{1DChiral,Gade,FC,GLL} while the clean spinless Hubbard model 
possesses only the Mott CDW instability. RG flow equations in related systems 
of spin-1/2 fermions were recently obtained independently in Ref.~\onlinecite{DellAnna}.

This work was supported in part by the NSF under Grant No.~DMR-00-75064 and by the UCSB Graduate 
Division (M.S.F.).


\begin{thebibliography}{99}
\bibitem{LeeRamakrishnan}For a review, see e.g. P. A. Lee, and T. V. Ramakrishnan, 
	Rev. Mod. Phys. \textbf{57}, 287 (1985).
\bibitem{1DChiral}
	For recent references, see e.g.:
	L. Balents and M. P. A. Fisher, Phys. Rev. B \textbf{56}, 12970 (1997); M. Bocquet, 
	Nucl. Phys. B \textbf{546}, 621 (1999).
\bibitem{Gade}R. Gade and F. Wegner, Nucl. Phys. B \textbf{360}, 213 
	(1991); R. Gade, Nucl. Phys. B \textbf{398}, 499 (1993).
\bibitem{Finkelstein}A. M. Finkel'stein, Zh. Eksp. Teor. Fiz. \textbf{84}, 
	168 (1983), Sov. Phys. JETP \textbf{57}, 97 (1983).
\bibitem{BK}For a review, see e.g. D. Belitz and T. R. Kirkpatrick, 
	Rev. Mod. Phys. \textbf{66}, 261 (1994).
\bibitem{2DMIT}For a recent account see:
	S. V. Kravchenko and M. P. Sarachik, {\it Metal-insulator
	transition in two-dimensional electron systems},
	Rep. Prog. Phys. {\bf 67}, 1 (2004);
	A. Punnoose and A. M. Finkel'stein, Science {\bf 310}, 289 (2005).
\bibitem{Auerbach}For a recent discussion, see e.g. A. Auerbach, \textit{Interacting 
	Electrons and Quantum Magnetism} (Springer-Verlag, New York, 1994).
\bibitem{Zirnbauer}M. R. Zirnbauer, J. Math. Phys. \textbf{37}, 4986 (1996).
\bibitem{ChiralOne} P. A. Lee and D. S. Fisher, Phys. Rev. Lett. 
	\textbf{47}, 882 (1981); A. Furusaki, Phys. Rev. Lett. \textbf{82}, 604
	(1998); M. Bocquet and J. T. Chalker, Phys. Rev. B 
	\textbf{67}, 054204 (2003); Ann. Henri Poincar\'e \textbf{4}, 
	S539 (2003).
\bibitem{GLL}S. Guruswamy, A. LeClair, and A. W. W. Ludwig, Nucl. Phys. B 
	\textbf{583}, 475 (2000). 
\bibitem{FC} M. Fabrizio and C. Castellani, Nucl. Phys. B \textbf{583}, 542 (2000);
\bibitem{ChiralTwo} 
	O. Motrunich, K. Damle, and D. A. Huse, Phys. Rev. B 
	\textbf{65}, 064206 (2002); C. Mudry, S. Ryu, and A. Furusaki, Phys. Rev. B \textbf{67}, 
	064202 (2003).	
\bibitem{BDIpaper}M. S. Foster and A. W. W. Ludwig, Phys. Rev. B \textbf{73}, 155104 (2006).
\bibitem{Hirsch}J. E. Hirsch, Phys. Rev. B \textbf{31}, 4403 (1985).
\bibitem{GSST}J. E. Gubernatis, D. J. Scalapino, R. L. Sugar, W. D. Toussaint, 
	Phys. Rev. B \textbf{32}, 103 (1985).
\bibitem{Shankar}R. Shankar, Rev. Mod. Phys. \textbf{66}, 129 (1994).
\bibitem{ClassAIII} Without interactions, the model is in class AIII 
	of Ref.~\onlinecite{Zirnbauer}.
\bibitem{Scalettar}M. Ulmke and R. T. Scalettar, Phys. Rev. B \textbf{55}, 4149 (1997);
	P. J. H. Denteneer, R. T. Scalettar, N. Trivedi, Phys. Rev. Lett. \textbf{87}, 
	146401 (2001).
\bibitem{ClassBDI} In the presence of time-reversal invariance (TRI), SLS is equivalent to the 
	usual particle-hole symmetry. 
\bibitem{AIIIlongpaper}M. S. Foster and A. W. W. Ludwig, (unpublished).
\bibitem{KELDYSH}M. L. Horbach and G. Schoen, Ann. Phys. (Leipzig) 
	\textbf{2}, 51 (1993); A. Kamenev and A. Andreev, Phys. Rev. B 
	\textbf{60}, 2218 (1999); C. Chamon, A. W. W. Ludwig, and C. 
	Nayak, Phys. Rev. B \textbf{60}, 2239 (1999). 	
\bibitem{OmegaMatrixDef}
	The symbol $\hat{\Omega}$ appearing in Eq.~(\ref{SD}) denotes a familiar matrix 
	(Refs.~\onlinecite{KELDYSH,AIIIlongpaper}) 
	diagonal in frequency (and Keldysh) indices.
\bibitem{AltshulerAronov}B. L. Altshuler and A. G. Aronov, in 
	\textit{Electron-Electron Interactions in Disordered Systems}, edited by 
	A. L. Efros and M. Pollak (North-Holland, Amsterdam, 1985.)
\bibitem{NPCS}E. P. Nakhmedov, V. Prigodin, S. \c Cal\i \c skan, 
	E. \c Sa\c s\i o\~glu, Phys. Rev. B \textbf{66}, 233105 (2002).
\bibitem{Aleiner}A. M. Rudin, I. L. Aleiner, and L. I. Glazman, Phys. Rev.
	B \textbf{55}, 9322 (1997); I. L. Aleiner, B. L. Altshuler, and M. E.
	Gershenson, Waves Random Media \textbf{9}, 201 (1999).
\bibitem{AleinerExplanation} Background density fluctuations become a source 
	of \emph{on-site} disorder in the presence of electron-electron interactions, 
	so that we may attribute this nontrivial correction to ``dynamic SLS breaking.''
\bibitem{NeggammasNote}
	We do not consider here $\gamma_{s} < 0$. 
\bibitem{DellAnna}
	L. Dell'Anna, Nucl. Phys. B {\bf 758}, 255 (2006).
\end{thebibliography}
\end{document}